\documentclass[conference]{IEEEtran}
\IEEEoverridecommandlockouts
\usepackage{cite}
\usepackage{amsmath,amssymb,amsfonts}
\usepackage{algorithmicx}
\usepackage{graphicx}
\usepackage{textcomp}
\usepackage{multicol, multirow}
\usepackage{booktabs}
\usepackage{xcolor}
\usepackage{algorithm}
\usepackage[algo2e,linesnumbered]{algorithm2e}
\usepackage{algpseudocode}
\usepackage{mathtools, nccmath}
\DeclarePairedDelimiter{\nint}\lfloor\rceil

\def\BibTeX{{\rm B\kern-.05em{\sc i\kern-.025em b}\kern-.08em
    T\kern-.1667em\lower.7ex\hbox{E}\kern-.125emX}}
\begin{document}

\title{Confidence-Aware Subject-to-Subject
Transfer Learning for Brain-Computer Interface\\

\thanks{
This work was partly supported by Institute of Information and Communications Technology Planning and Evaluation (IITP) grants funded by the Korea government (No. 2017-0-00451, Development of BCI based Brain and Cognitive Computing Technology for Recognizing User’s Intentions using Deep Learning; No. 2019-0-00079, Artificial Intelligence Graduate School Program(Korea University); No. 2021-0-00866, Development of BMI application technology based on multiple bio-signals for autonomous vehicle drivers).}
}


\author{
\IEEEauthorblockN{Dong-Kyun Han}
\IEEEauthorblockA{\textit{Dept. Brain and Cognitive Engineering} \\
\textit{Korea University}\\
Seoul, Republic of Korea \\
dk\_han@korea.ac.kr}
\\

\IEEEauthorblockN{Dong-Young Kim}
\IEEEauthorblockA{\textit{Dept. Artificial Intelligence}  \\
\textit{Korea University}\\
Seoul, Republic of Korea \\
dy\_kim@korea.ac.kr}
\and

\IEEEauthorblockN{Serkan Musellim}
\IEEEauthorblockA{\textit{Dept. Brain and Cognitive Engineering} \\
\textit{Korea University}\\
Seoul, Republic of Korea \\
serkanmusellim@korea.ac.kr}
\\

\IEEEauthorblockN{Ji-Hoon Jeong}
\IEEEauthorblockA{\textit{Dept. Artificial Intelligence}  \\
\textit{Korea University}\\
Seoul, Republic of Korea \\
jh\_jeong@korea.ac.kr}

}

\maketitle
\begin{abstract}
The inter/intra-subject variability of electroencephalography (EEG) makes the practical use of the brain-computer interface (BCI) difficult. 
In general, the BCI system requires a calibration procedure to tune the model every time the system is used. 
This problem is recognized as a major obstacle to BCI, and to overcome it, approaches based on transfer learning (TL) have recently emerged. 
However, many BCI paradigms are limited in that they consist of a structure that shows labels first and then measures ``imagery", the negative effects of source subjects containing data that do not contain control signals have been ignored in many cases of the subject-to-subject TL process.
The main purpose of this paper is to propose a method of excluding subjects that are expected to have a negative impact on subject-to-subject TL training, which generally uses data from as many subjects as possible.
In this paper, we proposed a BCI framework using only high-confidence subjects for TL training.
In our framework, a deep neural network selects useful subjects for the TL process and excludes noisy subjects, using a co-teaching algorithm based on the small-loss trick.
We experimented with leave-one-subject-out validation on two public datasets (2020 international BCI competition track 4 and OpenBMI dataset).
Our experimental results showed that confidence-aware TL, which selects subjects with small loss instances, improves the generalization performance of BCI.

\end{abstract}


\section{Introduction}
The brain-computer interface (BCI) is a technology that analyzes human brain signals to comprehend the user's intention and makes it possible for the user to communicate with external devices or systems via these signals \cite{lebedev2006brain,wolpaw2000brain}. 
For BCIs, a variety of brain imaging modalities are available \cite{zhang2017hybrid, mellinger2007meg},
with electroencephalography (EEG) being a well-known and commonly utilized brain signal\cite{chen2016high,jeong2020decoding}. Three types of paradigms are commonly utilized in EEG-based BCI: event-related potential (ERP) \cite{fazel2012p300,won2017motion,blankertz2011single,lee2018high} and steady-state visual potential (SSVEP) \cite{muller2005steady} based on stimulus presentation, and motor imagery (MI) \cite{pfurtscheller2001motor,channel,schirrmeister2017deep,jeong2020Brain} based on intuitive imagination. Recent studies have been proposed on paradigms like speech and visual imagery that emphasize increased intuitiveness\cite{lee2019towards}. 

One of the most difficult aspects of EEG analysis is that it changes over time and from subject to subject or session to session due to changes in the subject's physiological or psychological condition, as well as physical factors such as electrode positioning on the scalp.
The model's generalization performance on unseen data is hampered by these inter/intra-subject variability of EEG signals \cite{zhang2019strength}.
Therefore, general BCI systems require a calibration procedure to acquire subject/session-specific data to tune the model every time the system is used\cite{suk2011subject,lotte2010regularizing,lee2015subject}. 
This calibration procedure takes about 20-30 minutes of data collecting and model training time, which is inconvenient and ineffective. This calibration step must be simplified or removed in order for BCI to be used effectively.

Transfer learning (TL)-based techniques that leverage other subjects' data to help learn a limited quantity of calibration data or generate a generalized model have been developed for this purpose.
To address the aforementioned issue, TL approaches based on the subjects' or other subjects' previously obtained data have been investigated  \cite{kwon2019subject}.
In order to extract characteristics between subjects, a lot of research has been done. Depending on whether or not data from the target subject is used, it is separated into domain adaptation and domain generalization\cite{han2021domain}.
The majority of these studies use data from all subjects except the target subject to use as much data as possible.

Many contemporary BCI paradigms, on the other hand, show the target label first and collect data subsequently.
In reality, it's unclear if all participants (i.e. subjects) were able to correctly create a control signal by executing the required ``imagery" by focusing on all trials. As a result, there is a chance that samples/subjects with a negative impact are included in the TL process, which employs one's own or other subjects' data. 
Even in building a specific model, there are cases in which the performance deteriorates compared to the generalization model.
However, only a few studies have suggested ways to mitigate these negative effects \cite{jeon2021mutual, won2021selective}. 

In this paper, we proposed a confidence-aware subject-to-subject TL framework to mitigate the negative effect of noisy training samples which may doesn't contain control signals.
Here, we improved BCI systems generalization performance by excluding subjects with low-confidence samples during subject-to-subject TL training using subject selection based on the small loss trick.

\begin{figure}[t]
    \centering
    \includegraphics[width =0.8\columnwidth, height =7cm]{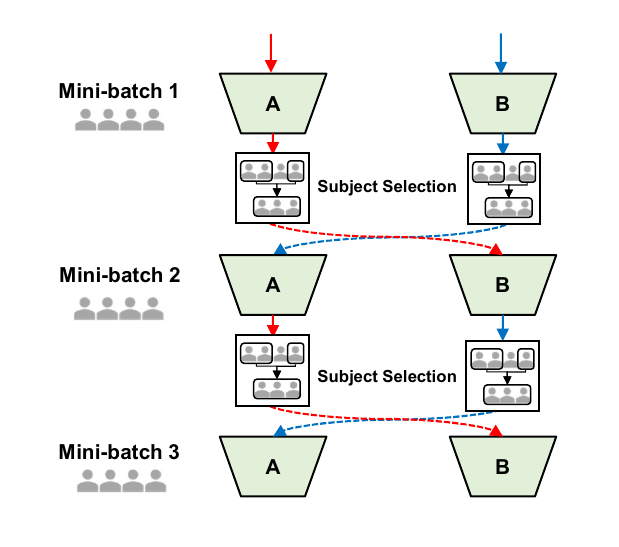}
    \caption{\label{fig:Fig.1.}
        Overview of error flow. The proposed method maintains two networks (A \& B) simultaneously, such as Co-teaching. In each mini-batch data, each network samples its small-loss subjects and teaches such small-loss subjects' loss instances to its peer network for the further training. Thus, the error flow displays the zigzag shape.}
\end{figure}

\begin{algorithm}[t]
1: {\bfseries Input} $w_f$ and $w_g$, learning rate $\eta$, fixed $\tau$, epoch $T_{k}$ and $T_{\max}$, iteration $M_{\max}$, subjects index $N$;

\For{$T = 1,2,\dots,T_{\max}$}{
	
	2: {\bfseries Shuffle} training set $\mathcal{D}$; \hfill //noisy dataset
	
	\For{$M = 1,\dots,M_{\max}$}
	{	
		3: {\bfseries Fetch} mini-batch $\bar{\mathcal{D}}$ from $\mathcal{D}$;
		
		4: {\bfseries Obtain} $\bar{N}_f = \arg\min_{N':|N'|\ge R(T)|N|}
		\Sigma{\ell(f, \bar{\mathcal{D}}_{N'})}$; 
	    
	    5: {\bfseries Obtain} $\bar{\mathcal{D}}_f = \{\bar{\mathcal{D}}_i\}, i \in  \bar{N}_f$ ; 
		
		\hfill //sample $R(T)\%$ small-loss subjects;
		
		6: {\bfseries Obtain} $\bar{N}_g = \arg\min_{N':|N'|\ge R(T)|N|}
		\Sigma{\ell(g, \bar{\mathcal{D}}_{N'})}$; 
		
		7: {\bfseries Obtain} $\bar{\mathcal{D}}_g = \{\bar{\mathcal{D}}_i\}, i \in  \bar{N}_g$ ; 
		
		\hfill //sample $R(T)\%$ small-loss subjects;
		
		8: {\bfseries Update} $w_f = w_f - \eta\nabla \ell(f,\bar{\mathcal{D}}_g)$; 
		
		\hfill //update $w_f$ by $\bar{\mathcal{D}}_g$;
		
		9: {\bfseries Update} $w_g = w_g - \eta\nabla \ell(g,\bar{\mathcal{D}}_f)$; 
		
		\hfill //update $w_g$ by $\bar{\mathcal{D}}_f$;
	}

	10: {\bfseries Update} $R(T) = 1 - \min\left\lbrace  \frac{T}{T_k} \tau, \tau \right\rbrace $;
}

11: {\bfseries Output $w_f$ and $w_g$}
\caption{Procedure of the proposed method.}
\label{alg:Co-teaching based subject selection}
\end{algorithm}

\section{Methods}

Our idea is to exclude subjects with low confidence during training. We proposed a BCI framework for cross-subject transfer learning while sorting confidence rankings and excluding a certain percentage of samples (Fig.\ref{fig:Fig.1.}).

\subsection{Co-teaching}
For confidence-aware training, samples with low confidence were excluded from the learning process using Co-teaching \cite{NEURIPS2018_a19744e2}.
Co-teaching was originally proposed to robustly train models on datasets with label noise.
Inspired by studies on the memorization effect of deep learning that neural networks first memorize training data with clean labels, Co-teaching try to select small-loss instances, and then use them to update the network robustly.
Further, to solve the problem of accumulated errors, Co-teaching trains two networks symmetrically and simultaneously. 
Each network filters noisy samples based on the so-called small loss trick. Then, it teaches the remaining small-loss samples to its peer network for updating the parameters.
\subsection{Confidence-aware Subject Selection}
In this section, we introduce how to select a subject with confidence in the cross-subject learning process illustrated in the Algorithm \ref{alg:Co-teaching based subject selection}, which consists of the fetching mini-batch step (step3) and the cross-update step (step 4-9). 

Inspired by the effect that the deep neural network learns the features of a prominent sample first, as mentioned above, we filtered samples during cross-subject training under the assumption that samples without a control signal will have a large loss. 
As with Co-teaching, we maintain two networks $f$ (with parameter $w_f$) and $g$ (with parameter $w_g$) and operate them in a mini-batch manner. However, it differs from general Co-teaching in that we use a dataset composed of multiple source subjects.

As with Co-teaching, we maintain two networks and operate them in mini-batch manners. 
Also, in applying sample filtering based on small loss, we use the sum of losses for each subject. Therefore, it is necessary to sample the mini-batch so that all subject's samples are evenly distributed in each mini-batch.
Let's denote training set composed $N$ source subjects $\mathcal{D} = $ ${\{\mathcal{D}_1, \mathcal{D}_2, \cdots, \mathcal{D}_N\}}$.
Then the extracted mini-batch is composed of $\bar{\mathcal{D}} = \{\bar{\mathcal{D}_1},\bar{\mathcal{D}_2}, \cdots,\bar{\mathcal{D}_N}\}$, where $i$-th subject's $\bar{\mathcal{D}_i}$ have $b$ samples. So each mini-batch has a total number of samples $B = b\times N$.

\begin{table*}[ht!]
\centering
\caption{The classification results (Balanced accuracy) of BCI competition dataset}
\label{table1}
\renewcommand{\arraystretch}{1}
\resizebox{\textwidth}{!}{
\begin{tabular}{llllllllllllllllll}
\hline
\multicolumn{1}{|l|}{\multirow{2}{*}{\textbf{Model}}} & \multicolumn{15}{c|}{\textbf{Subject}}                                                                                                                                                                                                                                                                                                                                                                                                                                                                                                                                                                                                                                   & \multicolumn{1}{l|}{\multirow{2}{*}{\textbf{Avg.}}} & \multicolumn{1}{l|}{\multirow{2}{*}{\textbf{Std.}}} \\ \cline{2-16}
\multicolumn{1}{|l|}{}                                & \multicolumn{1}{c|}{\textit{\textbf{1}}} & \multicolumn{1}{c|}{\textit{\textbf{2}}} & \multicolumn{1}{c|}{\textit{\textbf{3}}} & \multicolumn{1}{c|}{\textit{\textbf{4}}} & \multicolumn{1}{c|}{\textit{\textbf{5}}} & \multicolumn{1}{c|}{\textit{\textbf{6}}} & \multicolumn{1}{c|}{\textit{\textbf{7}}} & \multicolumn{1}{c|}{\textit{\textbf{8}}} & \multicolumn{1}{c|}{\textit{\textbf{9}}} & \multicolumn{1}{c|}{\textit{\textbf{10}}} & \multicolumn{1}{c|}{\textit{\textbf{11}}} & \multicolumn{1}{c|}{\textit{\textbf{12}}} & \multicolumn{1}{c|}{\textit{\textbf{13}}} & \multicolumn{1}{c|}{\textit{\textbf{14}}} & \multicolumn{1}{c|}{\textit{\textbf{15}}} & \multicolumn{1}{l|}{}                               & \multicolumn{1}{l|}{}                               \\ \hline
\multicolumn{1}{|l|}{DeepConvNet}                     & \multicolumn{1}{l|}{53.75}               & \multicolumn{1}{l|}{67.50}               & \multicolumn{1}{l|}{40.00}               & \multicolumn{1}{l|}{82.58}               & \multicolumn{1}{l|}{50.25}               & \multicolumn{1}{l|}{45.67}               & \multicolumn{1}{l|}{40.42}               & \multicolumn{1}{l|}{33.58}               & \multicolumn{1}{l|}{33.75}               & \multicolumn{1}{l|}{39.17}                & \multicolumn{1}{l|}{40.42}                & \multicolumn{1}{l|}{52.50}                & \multicolumn{1}{l|}{48.67}                & \multicolumn{1}{l|}{49.08}                & \multicolumn{1}{l|}{58.75}                & \multicolumn{1}{l|}{49.07}                          & \multicolumn{1}{l|}{12.67}                          \\ \hline
\multicolumn{1}{|l|}{EEGNet}                          & \multicolumn{1}{l|}{50.00}               & \multicolumn{1}{l|}{60.00}               & \multicolumn{1}{l|}{42.50}               & \multicolumn{1}{l|}{79.42}               & \multicolumn{1}{l|}{46.33}               & \multicolumn{1}{l|}{48.25}               & \multicolumn{1}{l|}{44.17}               & \multicolumn{1}{l|}{36.42}               & \multicolumn{1}{l|}{34.25}               & \multicolumn{1}{l|}{36.92}                & \multicolumn{1}{l|}{44.08}                & \multicolumn{1}{l|}{50.08}                & \multicolumn{1}{l|}{54.58}                & \multicolumn{1}{l|}{52.33}                & \multicolumn{1}{l|}{57.67}                & \multicolumn{1}{l|}{49.13}                          & \multicolumn{1}{l|}{10.91}                          \\ \hline
\multicolumn{1}{|l|}{ResNet1D}                        & \multicolumn{1}{l|}{51.67}               & \multicolumn{1}{l|}{61.25}               & \multicolumn{1}{l|}{48.25}               & \multicolumn{1}{l|}{83.42}               & \multicolumn{1}{l|}{51.08}               & \multicolumn{1}{l|}{48.42}               & \multicolumn{1}{l|}{39.67}               & \multicolumn{1}{l|}{32.92}               & \multicolumn{1}{l|}{31.17}               & \multicolumn{1}{l|}{39.08}                & \multicolumn{1}{l|}{38.33}                & \multicolumn{1}{l|}{51.25}                & \multicolumn{1}{l|}{56.00}                & \multicolumn{1}{l|}{55.83}                & \multicolumn{1}{l|}{57.17}                & \multicolumn{1}{l|}{49.70}                          & \multicolumn{1}{l|}{12.60}                          \\ \hline
\multicolumn{1}{|l|}{Ours}                            & \multicolumn{1}{l|}{54.58}               & \multicolumn{1}{l|}{61.42}               & \multicolumn{1}{l|}{48.17}               & \multicolumn{1}{l|}{84.75}               & \multicolumn{1}{l|}{48.00}               & \multicolumn{1}{l|}{44.67}               & \multicolumn{1}{l|}{42.17}               & \multicolumn{1}{l|}{38.58}               & \multicolumn{1}{l|}{34.83}               & \multicolumn{1}{l|}{41.42}                & \multicolumn{1}{l|}{39.08}                & \multicolumn{1}{l|}{50.08}                & \multicolumn{1}{l|}{60.08}                & \multicolumn{1}{l|}{61.08}                & \multicolumn{1}{l|}{57.00}                & \multicolumn{1}{l|}{\textbf{51.06}}                          & \multicolumn{1}{l|}{12.29}                          \\ \hline
                                                      &                                          &                                          &                                          &                                          &                                          &                                          &                                          &                                          &                                          &                                           &                                           &                                           &                                           &                                           &                                           &                                                     &                                                    
\end{tabular}
}
\end{table*}

When a mini-batch  $\bar{\mathcal{D}}$ of $B$ samples is formed (step3), we compute the sum of the loss per $N$ subjects from the mini-batch through network $f$ (resp. $g$).  
After that, $R(T)$ percentage of small-loss subjects out of the $N$ subjects are selected to become mini-batch $\bar{\mathcal{D}}_f$ (resp. $\bar{\mathcal{D}}_g$) (steps 4-7). Finally, the selected subjects' loss instances are fed into its peer network $g$ (resp. $f$) for parameter updates (steps 8 and 9). In step 10, we update $R(T)$, which controls how many subjects should be selected in each training epoch. We gradually
increase the drop rate, i.e., $R(T)$ becomes smaller, so that we can keep clean instances.

\begin{table}[bp]
    \caption{Specification of ResNet1D-18}
    \centering
    \begin{tabular}{ccc}
    \hline
    Block  & Output & ResNet1D-18    \\ \hline
    Input  & $E$×$T$ & -          \\ \hline
    Conv1  & 32×$\nint{T/2}$ & 7, 32, stride 2  \\ \hline \addlinespace[0.5ex]
    ResBlock1 & 32×$\nint{T/4}$  & $\begin{bmatrix} 3, 32 \\3, 32 \end{bmatrix}$×2, stride 2 \\ \addlinespace[0.5ex] \hline  \addlinespace[0.5ex]
    ResBlock2 & 64×$\nint{T/8}$  & $\begin{bmatrix} 3, 64 \\3, 64 \end{bmatrix}$×2, stride 2 \\ \addlinespace[0.5ex] \hline  \addlinespace[0.5ex]
    \multirow{2}{*}{ResBlock3} & \multirow{2}{*}{128×$\nint{T/64}$} & 4 max pool, stride 4  \\ \cline{3-3} \addlinespace[0.5ex]
        &  & $\begin{bmatrix} 3, 128 \\3, 128 \end{bmatrix}$×2, stride 2 \\ \addlinespace[0.5ex]  \hline \addlinespace[0.5ex]
    \multirow{2}{*}{ResBlock4} & \multirow{2}{*}{256×$\nint{T/512}$} & 4 max pool, stride 4  \\ \cline{3-3} \addlinespace[0.5ex]
        &  & $\begin{bmatrix} 3, 256 \\3, 256 \end{bmatrix}$×2, stride 2 \\ \addlinespace[0.5ex]  \hline \addlinespace[0.5ex]
        & 256×$\nint{T/512}$ & ELU, Adaptive average pool \\ \hline \addlinespace[0.5ex]
    Classification & $C$  & 2-d fc \\ \hline
    \end{tabular}
    \label{tab2}
\end{table}

\begin{table}[t]
\caption{The classification results of OpenBMI dataset}
\begin{center}
\centering
\begin{tabular}{|l|c|c|}
\hline
 \textbf{Model}       &  \textbf{Balanced accuracy} &  \textbf{Std.}  \\ \hline 
DeepConvNet & 84.24        & 7.15             \\ \hline
EEGNet      & 84.02        & 7.53             \\ \hline
ResNet1D   & 84.38        & 6.38             \\ \hline
Ours        & \textbf{85.07}        & 6.54             \\ \hline
\end{tabular}

\label{tab1}
\end{center}
\end{table}

\section{Experiments}
\subsection{Dataset}


Two publicly available datasets were used for this study, The two motor imagery datasets include:

\subsubsection{2020 international BCI competition track 4 dataset (http://brain.korea.ac.kr/bci2021/competition.php)} The dataset consists of 3-class grasp motor imagery tasks,  recorded for 3 sessions from 15 subjects, and labels from 2 sessions (Day 1 and Day 2) were released for competition. Each subject performed 50 trials per each grasping in one session (150 trials: 3 classes × 50 trials). We used data from the first and second sessions of 15 subjects where the label was released (4,500 trials). The data was sampled at 250Hz, and we used the raw data as it is.

\subsubsection{OpenBMI dataset \cite{Lee2019}}
The dataset consists of 2-class (right/left hand) motor imagery tasks,  recorded for 2 sessions from 54 subjects. Each subject performed 100 trials per each class in one session (200 trials: 2 classes × 100 trials). We used data from the both sessions of 54 subjects (21,600 trials). The data was sampled at 1000Hz, and we downsampled it to 250 Hz.

\subsection{Data Configuration}
We assumed that trials without control signals would contain features similar to ``rest" class, and we tried to remove noisy data by reflecting these assumptions in the learning process.
However, since neither of the datasets include independent ``rest" trials, we added to the dataset by segmenting the rest interval measured just before each imagery trial. Therefore, the number of trials used in the experiments was doubled and the number of classes increased by one. All data were segmented into 3 seconds after onset to fit all trials in the rest time period. 
Hence, the shape of the input sample is 750 (3sec)$\times$the number of electrodes, and in the case of competition dataset, the number of electrodes was 60, and in the case of OpenBMI dataset, it was 62.

We conducted experiments by leave-one-subject-out validation.
For the competition dataset, for example, the samples excluding one subject from the 15 subjects' samples (i.e. 14 subjects) were used as the source data to be used for cross-subject training, and the excluded data was used as the test data. 
We split the source data set at a ratio of 9:1 and used them as train data and validation data, respectively. 

\subsection{Experimental Details}
We used ResNet1D with 18 layers (ResNet1D)\cite{han2021domain}, as our backbone network, which recently achieved outstanding performance in a large motor imagery dataset, and mental state classification. 
ResNet1D consists of a convolutional layer, followed by ResNet building blocks (ResBlocks), max pooling layers, and ELU activation.
ResBlock consists of ELU activations and one-dimensional convolutional layers.
Table \ref{tab2} shows the detail of ResNet1D, $T$ denotes the time period of the input data, $E$ denotes the number of electrodes in the input data, and $C$ denotes the number of classes.

For comparison, we experimented three baseline classifiers: DeepConvNet \cite{schirrmeister2017deep}, EEGNet \cite{lawhern2018eegnet}, and ResNet1D.
DeepConvNet is a convolutional neural network (CNN) structure that consists of 4 convolution-pooling blocks in which the first convolution-pooling block is a combination of temporal and spatial convolution filters. 
EEGNet is a shallow CNN structure compared to DeepConvNet, which uses depth-wise separable convolution filters instead of convolution-pooling blocks.

The baseline classifiers were trained only with the cross-entropy loss. We evaluated our experiments with balanced accuracy because of the imbalanced dataset. For evaluation, we used the model with the highest balanced accuracy in the validation set.

The loss function was optimized by the Adam algorithm with a learning rate of 0.01. We also used the cosine annealing learning rate scheduler \cite{loshchilov2016sgdr}. The subject wise batch size $b$ for training was 8. We set the hyperparameters in $R(T)$, with $T_k = 10$ and $\tau = 0.2$ for competition dataset and $T_k = 10$ and $\tau = 0.1$ for OpenBMI dataset.

\section{Results and Conclusion}
Table \ref{table1} shows the classification results using the 2020 international BCI competition track 4 dataset.
Our proposed method achieved the highest performance (51.06\%) in terms of average balanced accuracy of all subjects. The performances of the other models were 49.07\%, 49.13\%, and 49.70\% for DeepConvNet, EEGNet, and ResNet1D, respectively.
Table \ref{tab1} shows the classification results using the OpenBMI dataset.
Our proposed method also achieved the highest performance (85.07\%). The performances of the other models were 84.24\%, 84.02\%, and 84.38\% for DeepConvNet, EEGNet,and ResNet1D, respectively.

In this paper, we proposed a confidence-aware subject-to-subject TL framework. In our framework, the deep learning model excludes subjects with low confidence during training. Based on the experiments, applying Co-teaching like confidence-aware subject selection showed an improvement in terms of generalization performance. 

 We only added the ``rest" class and evaluated the classification performance, but in future work, it is necessary to analyze the excluded samples and find out what kind of label noise may exist in BCI data and what class it is actually associated with and what kind of label noise it corresponds to. Also, neurophysiological analysis of these noisy samples is required.

\bibliographystyle{IEEEtran}
\bibliography{refs}

\vspace{12pt}

\end{document}